%% file: main.tex
\documentclass[ALICE,manyauthors]{cernphprep}
\usepackage[comma,square,numbers,sort&compress]{natbib}
\usepackage{hyperref}
\usepackage{lineno}
\usepackage{xspace}
\usepackage[T1]{fontenc}
\usepackage{orcidlink}
\usepackage[dvipsnames]{xcolor}
\input{commands.tex}
\begin{document}

\begin{titlepage}
\PHyear{2026}       
\PHnumber{176}      
\PHdate{12 June}  

\title{Energy-energy correlators in p--Pb collisions at $\sqrt{s_{\rm NN}}=5.02$ TeV}
\ShortTitle{Energy-energy correlators in p--Pb collisions at $\sqrt{s_{\rm NN}}=5.02$ TeV}   

\Collaboration{ALICE Collaboration\thanks{See Appendix~\ref{app:collab} for the list of collaboration members}}
\ShortAuthor{ALICE Collaboration} 

\begin{abstract}

This paper presents the first measurement of the two-point energy–energy correlator (EEC) inside charged-particle jets in \pPb collisions at \snn = 5.02 \TeV with the ALICE experiment. The two-point EEC, an infrared- and collinear-safe measure of angular energy flow within jets, is used to probe both perturbative and non‑perturbative stages of the jet evolution. The EEC is reported in 20–80 \GeVc jets in \pPb collisions and compared to a measurement of the EEC in \pp events at $\sqrt{s} = 5.02$ TeV. A modification is observed in the 20–40 \GeVc interval, with an enhancement at large opening angles and a suppression at small angles relative to pp collisions. The dependence of this modification on jet transverse momentum, rapidity, and forward event activity is investigated. These results provide new constraints on cold nuclear matter effects relevant for disentangling initial‑ and final‑state contributions to jet-structure modifications. An understanding of these cold nuclear matter effects is also relevant for interpreting EEC measurements in heavy‑ion collisions.

\end{abstract}
\end{titlepage}

\setcounter{page}{2} 


\input{Ch1_Introduction}

\input{Ch2_Experiment}

\input{Ch3_Analysis}
\input{Ch4_Systematics}

\input{Ch5_Results}
\input{Ch6_Conclusions}


\newenvironment{acknowledgement}{\relax}{\relax}
\begin{acknowledgement}
\section*{Acknowledgements}
\input{fa_2026-06-03_Opt_C.tex}
\end{acknowledgement}

\bibliographystyle{utphys}   
\bibliography{bibliography}

\newpage
\appendix

%
%

\section{The ALICE Collaboration}
\label{app:collab}
\input{Alice_Authorlist_2026-06-03_Opt_C.tex}  
\end{document}

%% file: commands.tex
\newcommand{\pp}           {pp\xspace}

\newcommand{\PbPb}         {\mbox{Pb--Pb}\xspace}
\newcommand{\pA}           {\mbox{pA}\xspace}
\newcommand{\pPb}          {\mbox{p--Pb}\xspace}

\newcommand{\s}            {\ensuremath{\sqrt{s}}\xspace}
\newcommand{\snn}          {\ensuremath{\sqrt{s_{\mathrm{NN}}}}\xspace}
\newcommand{\pt}           {\ensuremath{p_{\rm T}}\xspace}
\newcommand{\kt}           {\ensuremath{k_{\rm T}}\xspace}
\newcommand{\jpt}           {\ensuremath{p_{\rm T}^{\rm ch\> jet}}\xspace}
\newcommand{\pttrk}         {\ensuremath{p_{\rm T, track}}\xspace}
\newcommand{\rl}           {\ensuremath{R_{\rm L}}\xspace}
\newcommand{\ptrl}         {\ensuremath{\langle p_{\rm T}^{\rm ch\> jet}\rangle R_{\rm L}}\xspace}

\newcommand{\etarange}[1]  {\mbox{$\left | \eta \right |~<~#1$}}
\newcommand{\jetarange}[1]  {\mbox{$\left | \eta_{\rm jet} \right |~<~#1$}}

\newcommand{\RpPb}         {\ensuremath{R_{\rm pPb}}\xspace}
\newcommand{\EEC}          {\ensuremath{\Sigma_{\rm EEC}}\xspace}

\newcommand{\nineH}        {$\sqrt{s}~=~0.9$~Te\kern-.1emV\xspace}
\newcommand{\seven}        {$\sqrt{s}~=~7$~Te\kern-.1emV\xspace}
\newcommand{\twoH}         {$\sqrt{s}~=~0.2$~Te\kern-.1emV\xspace}
\newcommand{\twosevensix}  {$\sqrt{s}~=~2.76$~Te\kern-.1emV\xspace}
\newcommand{\five}         {$\sqrt{s}~=~5.02$~Te\kern-.1emV\xspace}
\newcommand{\twosevensixnn}{$\sqrt{s_{\mathrm{NN}}}~=~2.76$~Te\kern-.1emV\xspace}
\newcommand{\fivenn}       {$\sqrt{s_{\mathrm{NN}}}~=~5.02$~Te\kern-.1emV\xspace}

\newcommand{\GeVc}         {Ge\kern-.1emV/$c$\xspace}
\newcommand{\MeVc}         {Me\kern-.1emV/$c$\xspace}
\newcommand{\TeV}          {Te\kern-.1emV\xspace}
\newcommand{\GeV}          {Ge\kern-.1emV\xspace}
\newcommand{\MeV}          {Me\kern-.1emV\xspace}
\newcommand{\GeVmass}      {Ge\kern-.2emV/$c^2$\xspace}
\newcommand{\MeVmass}      {Me\kern-.2emV/$c^2$\xspace}

\newcommand{\ITS}          {\rm{ITS}\xspace}

\newcommand{\SPD}          {\rm{SPD}\xspace}

\newcommand{\TPC}          {\rm{TPC}\xspace}

\newcommand{\VZEROA}       {\rm{V0A}\xspace}
\newcommand{\VZEROC}       {\rm{V0C}\xspace}

\definecolor{red}{HTML}{ff0037}

%% file: Ch1_Introduction.tex
\section{Introduction}

 In high-energy hadronic collisions, showers of secondary partons (quarks and gluons) are produced in hard scattering processes between the quarks and gluons of the colliding particles. The high transverse-momentum (\pt) partons from the initial scattering undergo quantum chromodynamic (QCD) radiation, splitting into lower-energy partons, until the hadronization energy scale is achieved. Below this scale, partons undergo confinement into color-neutral hadrons, which ultimately stream into detectors. The collimated sprays of hadrons, called jets, provide a clean probe of perturbative and non-perturbative QCD dynamics. Their internal structure, called jet substructure, encodes the pattern of parton branching and hadronization~\cite{salam-jetography, larkoski-sub-rev}.

The energy–energy correlator (EEC) quantifies angular correlations of energy flow within jets, providing direct sensitivity to both early-time perturbative splittings and late-time hadronization dynamics~\cite{cheng-moult-zhang-zhu-2020, opencmsdata, lee-mecaj-moult-2025}. The EEC was originally proposed as an event-shape observable for $e^+e^-$ collisions~\cite{ep-eec-basham, opal-as-lep, opal-as-lep-2, eec-ep-kardos}, and was also suggested as a tool for constructing jet discriminators~\cite{jet-discr-grigoriev, SVESHNIKOV1996403, GRIGORIEV200342}. It is infrared- and collinear-safe (IRC-safe)~\cite{irc-safe} and, for vacuum collisions, can be perturbatively calculated from first principles in QCD~\cite{eec-ep-kardos}. EEC observables offer precision probes of perturbative and non-perturbative QCD dynamics across a wide range of collision types~\cite{moult2025energycorrelatorsjourneytheory}.

At ALICE, the EEC is experimentally defined as a \pt-weighted two-particle correlation as a function of \rl, the angular distance between pairs of particles:
\begin{equation}
    \Sigma_{\rm EEC}(\rl) = \frac{1}{N_{\rm jet}\cdot\Delta} \int_{\rl-\frac{1}{2}\Delta}^{\rl+\frac{1}{2}\Delta}\sum_{\rm jets}\sum_{i,j}\frac{p_{{\rm T},i}p_{{\rm T},j}}{(\pt^{\rm jet})^2}\delta(\rl'-R_{{\rm L},ij})d\rl'.
\end{equation}
This defines the two-point EEC. The sum runs over all final state charged hadron pairs $(i, j)$ inside the jet. The angular distance between the two tracks in each pair in the $\eta-\varphi$ (rapidity-azimuth) plane is defined as $R_{{\rm L}, ij} = \sqrt{(\varphi_j-\varphi_i)^2 + (\eta_j-\eta_i)^2}$. The angular bin width is denoted as $\Delta$, and $N_{\rm jet}$ is the total number of jets. This definition can easily be extended to higher-order correlators by summing over sets of three or more particles in the jet, and choosing a suitable characteristic angular distance~\cite{opencmsdata}.

The $p_T$ weight suppresses soft radiation, reducing the EEC's sensitivity to higher-order corrections, the underlying event, and soft jet fragments -- making it particularly achievable to calculate theoretically. Theoretical developments have enabled the calculation of the EEC in proton-proton (\pp) collisions to high accuracy~\cite{cheng-moult-zhang-zhu-2020, e3c-ratio-theory}. Various theoretical predictions exist for the EEC in jets produced in lead-lead (\PbPb) collisions, where a host of other modifications come into effect~\cite{Andres_2023, andres-pbpb-eec, andres2024, andres-selection-bias, andres2025, barata-caucal-2025, barata-milhano-2025, xnw-pbpb-eec, hybrid-pbpb-eec, andres2024-pA, barata2025}. 

The EEC has been previously reported in \pp and \PbPb collisions by ALICE, ATLAS, and CMS at the LHC~\cite{alice-pp-eec, ATLAS:2023eec, cms-pp-eec, cms-pbpb-eec}, and in \pp collisions by STAR at RHIC~\cite{star-eec}. The \pp measurements~\cite{alice-pp-eec, cms-pp-eec, star-eec} show that the perturbative and non-perturbative phases of jet evolution populate different angular regions. Additionally, the strong coupling constant $\alpha_s$ can be extracted via ratios of the three-point energy correlator (EEEC) to the two-point correlator (EEC) in \pp collisions~\cite{ATLAS:2023eec, cms-pp-eec}, as proposed in~\cite{e3c-ratio-theory}. The \PbPb measurements by CMS~\cite{cms-pbpb-eec} indicate that the presence of the quark–gluon plasma (QGP) leads to modifications in the large $R_L$ region of the EEC.

This work presents the first measurement of the EEC inside jets produced in \pPb collisions at \snn = 5.02 \TeV at the LHC. Comparing jet substructure between \pp and proton-nucleus (\pA) collisions 
provides insights into the effects of a cold (non-deconfined) QCD medium surrounding the colliding nucleons~\cite{Connors:2017ptx, grosseoetringhaus2024decadecollectivitysmallsystems, cnm-jet-predictions, alice-qcd-summary}. Multiple scatterings in the nucleus or interactions between cold nuclear matter (CNM) and the early parton shower can make an imprint on jet-substructure observables. However, jet measurements in \pPb collisions generally show small or no modifications relative to \pp collisions. Observables such as the jet nuclear modification factor (\RpPb)~\cite{ppb-rpa}, the jet fragmentation transverse momentum ($j_{\rm T}$) distributions~\cite{alice-jt}, the jet mass~\cite{alice-jet-mass}, the fragmentation function~\cite{atlas-ppb-frag, cms-ppb-frag}, and the dijet $k_{\rm T}$~\cite{dijet-kt} do not indicate sizable transverse-momentum broadening or strong final-state energy loss. These results, generally probing jets down to 40–60 \GeVc, are consistent with expectations that jets traverse a relatively dilute cold nuclear environment where only modest initial-state effects (shadowing, nuclear PDFs) or weak multiple scattering may occur. Deviations from the \pp baseline have been reported for isolated prompt-photon production in \pPb events, and for jet fragmentation in high-multiplicity \pp events~\cite{alice-measurementisolatedpromptphoton, hmppzch}. Possible explanations for these modifications include gluon shadowing in the lead nucleus and multiparton interactions, respectively.

In a cold nuclear medium, soft multiple scattering or modest transverse kicks could alter the angular distribution of energy flow, affecting the large-angle region of the EEC. Several theoretical frameworks propose possible origins, including transverse-momentum broadening, interactions with partonic comovers, twist-4 Operator Product Expansion (OPE) corrections, and Glauber gluon interactions~\cite{zhongbo-eec, berndt-eec, andres2024-pA, weiyao-ppb}. 

This paper presents the first results of EEC measurements in \pPb collisions at the LHC. The ALICE detector's excellent tracking capabilities~\cite{alice-qcd-summary} allow for the construction of the EEC using charged-particle tracks inside charged-particle jets. The measurement is reported in charged-particle jet \pt (\jpt) intervals of 20–40, 40–60, and 60–80 \GeVc. These results are compared with state-of-the-art Monte Carlo models and theory calculations. The dependence of the EEC on the \pt of the jet's constituents, on the jet's pseudorapidity, and on the \pPb event activity is also investigated. The transition region between the perturbative and non-perturbative regimes, visible as a peak in the EEC distributions, is also analyzed to study hadronization as a function of jet \pt.

%% file: Ch2_Experiment.tex
\section{Experimental setup and datasets}

ALICE is a general-purpose heavy-ion experiment at the LHC. This analysis utilizes the central-barrel tracking system to measure jets and forward/backward detectors for triggering and to determine event activity. The central tracker consists of the Inner Tracking System (\ITS)~\cite{alice-its} and the Time Projection Chamber (\TPC)~\cite{alice-tpc}. These sit inside a 0.5 T magnetic field and have a high efficiency for detection of charged particles with \pt ($p_{\rm T, track}$) between 0.15 and 100 \GeVc, at pseudorapidity $\etarange 0.9$, and azimuthal angle $0 < \varphi < 2\pi$. The single-track efficiency is \pt dependent, increasing from $\sim$60\% at 0.15 \GeVc to $\sim$80\% at 1 \GeVc, and remaining above $\sim75$\% at higher \pt. The momentum resolution, $\sigma_p/p$, is $\approx1$\% at $p_{\rm T, track}$ = 1 \GeVc, increasing to $\approx4$\% at $p_{\rm T, track}$ = 4 \GeVc~\cite{alice-pp-eec}. The angular resolution is $\approx$1 mrad at $p_{\rm T, track}$ = 1 \GeVc, decreasing to $<$ 0.6 mrad for $p_{\rm T, track} >$ 5 \GeVc~\cite{charm-eec}. More detailed descriptions of the ALICE detector can be found in Refs.~\cite{alice-performance, alice-joi}. 

The results presented here are from LHC Run 2 data; \pp data were collected at \s = 5.02 \TeV in 2017, and \pPb data were collected at \snn = 5.02 TeV in 2016. Both data samples were collected with a minimum bias (MB) trigger, requiring coincident hits in the V0 detector, which is composed of two arrays of scintillator counters (\VZEROA, \VZEROC) at forward and backward rapidity~\cite{alice-v0}. The V0 arrays were located on either side of the nominal interaction point, covering the full azimuth within the ranges $2.8 < \eta < 5.1$ for \VZEROA, and $-3.7 < \eta < -1.7$ for \VZEROC. The amplitude in the \VZEROA, located in the Pb-going direction, was also used to determine the event activity of the collisions and is frequently used as a centrality estimator in \pPb measurements~\cite{centrality-ppb}. In this analysis, the \VZEROA amplitude is used to define two event activity classes (high-/low-multiplicity) for \pPb events.

Events were required to have a primary vertex, reconstructed using the Silicon Pixel Detector (\SPD), within 10 cm of the nominal interaction point to provide uniform acceptance and high tracking efficiency. Events with more than one reconstructed collision vertex were removed to avoid pileup. In total, 870 million MB \pp events passed the event selection criteria, corresponding to an integrated luminosity of 18.0 $\pm$ 0.4 nb$^{-1}$~\cite{ALICE-pp-lumi}. For the \pPb data, 624 million events passed these criteria, corresponding to an integrated luminosity of 298 $\pm$ 11 $\mu$b$^{-1}$~\cite{ALICE-ppb-lumi}. Because of the asymmetry of the colliding beam energies (protons at 4 \TeV, and fully stripped $^{208}_{82}$Pb ions at 1.58 \TeV per nucleon~\cite{ALICE-ppb-lumi}), a rapidity shift $\Delta y$ = 0.465 is needed in the direction of the proton beam to transform from the ALICE laboratory frame to the nucleon–nucleon center-of-mass frame.

%% file: Ch3_Analysis.tex
\section{Analysis Methods}

\subsection{Jet selection}

Charged particles measured with the central-barrel tracking system are selected using the same criteria as in Refs.~\cite{alice-performance, alice-jet-quenching}, including a \pt cutoff of 0.15 \GeVc. The selected charged tracks are clustered into jets using the anti-\kt algorithm~\cite{antikt} as implemented in the FastJet 3.4.0 package, with resolution parameter $R=0.4$~\cite{fastjet}. The E-scheme recombination is used, assigning to all charged particles the pion mass. Reconstructed jets with \jetarange{0.9}$-R$ are used to avoid detector edge effects in the jet reconstruction. 

In both \pp and \pPb collisions, reconstructed jets from the hard scattering are contaminated by soft background that does not originate from the hard process, called the underlying event (UE). Background particles enter the EEC in two places: in the jet \pt, which is inflated by the presence of UE particles, and in the EEC distribution itself, which has a combinatorial background introduced by pairs containing one or two UE particles. 

To address background contributions to the jet \pt, the \jpt of selected \pp and \pPb signal jets was corrected for the average UE contribution~\cite{rhosub} using the formula
\begin{equation}
    \jpt = p_{\rm T}^{\rm ch\> jet, raw} - \rho_{\rm ch}\cdot A_{\rm jet},
    \label{eq-jptcorr}
\end{equation}
where the transverse-momentum density $\rho_{\rm ch}$ of the UE was estimated on an event-by-event basis~\cite{rhosub-2}:
\begin{equation}
    \rho_{\rm ch} = \textrm{median} \left\{ \frac{p_{\rm T, jet}^{k_{\rm T}}}{A_{\rm jet}^{k_{\rm T}}} \right\}\cdot C.
\end{equation}
Here, $p_{\rm T, jet}^{k_{\rm T}}$ and $A_{\rm jet}^{k_{\rm T}}$ are the transverse momentum and area of background jet-like objects found in the same event using the \kt algorithm. The \kt algorithm for this purpose uses $R=0.4$ and \jetarange{0.9}. The active area~\cite{jet-active-area} is estimated by distributing ghost particles with infinitesimal \pt into the $\eta-\varphi$ acceptance. The two highest-\pt jets are excluded from the background estimation to discount the impact of physical jets from $\rho_{\rm ch}$~\cite{Cacciari_2011}. The scaling factor $C$ is used to account for regions without particles. It is defined as
\begin{equation}
    C = \frac{\sum_{\rm j}A_{\rm j}}{A_{\rm acc}},
\end{equation}
where $A_{\rm j}$ is the area of each \kt jet that is not a pure ghost jet, and $A_{\rm acc}$ is the area of charged-particle acceptance.

\subsection{EEC background subtraction}
\label{jet-sel}

To reduce the impact of non-jet particles from the UE, only constituents inside the reconstructed jets with \pttrk $> 1$ \GeVc are selected to enter the EEC. This 1 \GeVc threshold is varied during the analysis; results are shown in section~\ref{sec:trackpt}. Some UE particles will still remain in the jet, so the EEC distribution itself has to be corrected for contributions from track pairs containing one or two background tracks. These are hereafter referred to as signal-background and background-background pairs. 

A perpendicular cone method is used to estimate these contributions~\cite{ALICE-perp-cone}. Consider a circular cone with a radius equal to the signal jet's resolution parameter, located at the same $\eta$ but perpendicular in $\varphi$: $\Delta \varphi = \pm$90$^\circ$. For each signal jet, there are two such perpendicular cones. The UE is the predominant contribution to the tracks in these cones, which are located in the transverse region away from the jet. 

The tracks in the perpendicular cones can be treated as a proxy for the UE tracks under the signal jet if the UE in \pPb events is azimuthally symmetric. Measurements of elliptic flow indicate the degree of azimuthal anisotropy. In \pPb collisions, the integrated $v_2$ is $\sim 0.06$, roughly the same value as in 13 TeV \pp events~\cite{ppb-flow}. In 0-5\% central \pPb events, $v_2$ is comparable to very peripheral (70–80\% central) \PbPb events~\cite{ALICE-v2-ppb}. This suggests that the degree of asymmetry in \pPb events is small, so the UE is assumed to be nearly azimuthally symmetric.

To estimate the background-background and signal-background correlations, the tracks from one perpendicular cone are rotated by $\mp90^\circ$ so they lie on top of the signal jet, giving a new combined jet. The tracks in the combined jet are tagged either as tracks from the original jet (``jet'' tracks) or tracks from the perpendicular cone (``perp'' tracks). Pairs can now be constructed from the combined jet. This is performed separately for each of the two $\pm$90$^\circ$ perpendicular cones. 

These pairs fall into one of three types: jet-jet pairs, jet-perp pairs, or perp-perp pairs. The \EEC obtained from perp-perp pairs provides a good description of the background-background contribution, as it captures true correlations between nearby background particles. Similarly, the \EEC from jet-perp pairs describes the signal-background contribution. However, the jet-perp \EEC can also include background-background-like pairs, where one UE track from the jet cone has been paired with a UE track from the perpendicular cone. These pairs do not describe correlations that would be present within the jet cone. To remove this latter part of the jet-perp distribution, a similar type of uncorrelated pair is constructed by using one track from each of the two perpendicular cones, hereafter referred to as perp-perp$'$ pairs. The overall estimate of the background level (the combined signal-background and background-background contribution) can be expressed as
\begin{equation}
    \EEC(\mathrm{background}) = \EEC(\mathrm{jet}\text{-}\mathrm{perp}) + \EEC(\mathrm{perp}\text{-}\mathrm{perp}) - \EEC(\mathrm{perp}\text{-}\mathrm{perp}').
    \label{bkgtoteq}
\end{equation}
Since there are two perpendicular cones ($\pm90^\circ$) for each jet, the average of the two jet-perp and perp-perp EECs is used in the rule defined above. For each background pair (jet-perp, perp-perp, or perp-perp$'$ pair), the energy weight is taken to be $p_{{\rm T},1}p_{{\rm T},2}/(p^{\textrm{ch jet}}_{{\rm T}})^2$ regardless of the origin of the two tracks in the pair. The \jpt in the energy weight denominator is the corrected jet \pt defined in Eq.~\ref{eq-jptcorr}. A very similar background subtraction technique was implemented in the CMS \PbPb EEC measurement, using mixed events~\cite{cms-pbpb-eec}.

 The background obtained following the rule in Eq.~\ref{bkgtoteq} is subtracted from the jet-jet EEC to arrive at a background-corrected signal EEC distribution. Neither background distribution ever rises above a 10\% level, and furthermore at low $R_L$, the background is a $\sim$1\% effect.

\subsection{Detector effects}
Detector effects are corrected using \pPb collisions simulated with DPMJET~\cite{dpmjet} and propagated through a GEANT3 model~\cite{geant} of the ALICE detector. DPMJET is a multipurpose Monte Carlo event generator that uses the Dual Parton Model (DPM)~\cite{dpm} to describe soft interactions for hadron-hadron, hadron-nucleus, and nucleus-nucleus interactions. It uses the Glauber-Gribov multiple scattering formalism, and includes models for photoproduction, diffractive processes, the intranuclear cascade, and soft hadron-nucleus interactions. The corrections for the \pp baseline were performed using collisions simulated with PYTHIA 8 (version 8.210)~\cite{pythia8} Monash tune~\cite{pythia-monash}. 

The dominant detector effects considered in this measurement are the track-momentum resolution, angular resolution, and efficiency for both single tracks and pairs of tracks. The track-momentum resolution and efficiency affect the calculated energy weights, the number of reconstructed pairs, and the reconstructed \jpt. Track pairs with very small angular separation have very similar trajectories, which can cause one or both of the tracks to be missed or reconstructed incorrectly. This is called the track-merging effect, and was quantified in ALICE femtoscopy measurements~\cite{track-merging}. Track merging decreases the pair efficiency for tracks at small \rl, causing a reduced $\Sigma_{\rm EEC}$ in that region. To limit the impact of this effect, results are reported for \rl $> 0.01$. The magnitude of the corrections is below 10\%, except for the \rl bins near 0.01, where the correction is around 30\%. 

Detector effects are typically corrected using an unfolding procedure. However, in this analysis, the excellent momentum and position resolution of the ALICE detectors result in little bin migration in \rl. Hence, as in the ALICE \pp measurement~\cite{alice-pp-eec}, a two-dimensional bin-by-bin correction is used instead of unfolding. The correction factors are obtained by comparing the generator-level and detector-level information in simulation for each \jpt and \rl interval. This method can be sensitive to the simulation event generator, so a systematic uncertainty is included to account for this (described in Sec. 4). A cross-check was performed using a three-dimensional Bayesian unfolding technique in \rl and \jpt, with the pairwise energy weight treated as an additional axis~\cite{DAgostini:1994fjx, Adye:2011gm}. The difference between the unfolded and bin-by-bin corrected distributions is included in the systematic uncertainties as well.

%% file: Ch4_Systematics.tex
\section{Systematic uncertainties}

The main sources of systematic uncertainty for this analysis are the single-track efficiency, pair efficiency, background subtraction, detector-effect correction method, and event-generator dependence of the correction factors.

The single-track efficiency uncertainty is estimated to be $\sim$3\% in \pp events based on previous studies, where the track selection and ITS-TPC matching criteria were varied~\cite{trkeff-syst}. In p–Pb collisions, the value was found to increase with \pt from 1\% at
low \pt ($\sim$0.5 \GeVc) up to 2.5\% at high \pt ($\sim$14 GeV/c)~\cite{ppb-rpa}. This systematic uncertainty is assessed for the \pPb EEC by randomly rejecting 1–2.5\% of the detector-level tracks in simulated events in a \pt-dependent manner, and recalculating the correction factors. For the \pp baseline, 3\% of detector-level tracks in simulated events are randomly rejected and the correction factors are recalculated. In both \pPb and \pp events, an \rl-dependent uncertainty is quoted for this, which includes both the direct effect of track-efficiency variations and their propagated impact on the reconstructed \jpt.

The pair-efficiency uncertainty is evaluated by studying how the corrected EEC varies under different pair-selection criteria. Selections are made on ${\rm d}\eta=\lvert\eta_1-\eta_2\rvert$ and ${\rm d}\varphi^*=\lvert\varphi^*_1-\varphi^*_2\rvert$, where $\varphi^*$ is the azimuthal angle of a track in the laboratory frame at the outermost edge of the ALICE TPC, at a radial distance of 1.6m. In \pp and \pPb events, pairs with ${\rm d}\eta < 0.008$ and ${\rm d}\varphi^* < 0.01$ are rejected in data and in simulated events. In the nominal result no such cuts are made. The resulting difference in the corrected EEC is $\sim2$\% at small \rl and negligible at large \rl. This is used as an \rl-dependent uncertainty.

The uncertainty from the background subtraction is determined by varying the size of the perpendicular cone. Instead of using a fixed cone radius of $R=0.4$, the perpendicular cone radius is chosen to match the jet area and $R$ is taken to be $\sqrt{A_{\rm jet}/\pi}$. The subtracted EEC where the background is determined with a fixed radius is compared to the background obtained with this variable radius in both \pp and \pPb events. Additionally, in \pPb events the subtracted EEC using just the $+90^\circ$ cone and the subtracted EEC using just the $-90^\circ$ cone are compared. This provides a systematic uncertainty related to the assumption of azimuthal symmetry in \pPb events. Furthermore, in \pp events, a systematic uncertainty is assigned to the area-based \jpt correction to account for the sparsity of the events and possible resultant biases in the background estimation. To estimate this uncertainty, the \jpt was reduced by the average $\rho A_\mathrm{jet}$ in PYTHIA simulations and compared to the unsubtracted EEC. This systematic uncertainty was determined to be 3\%, independent of \rl.  With these methods, an \rl-dependent uncertainty for the total background-subtraction process is obtained in \pPb and \pp events. It is generally $\sim3\%$, but rises to $\sim7\%$ in \pPb events at large \rl in the 40–60 \GeVc and 60–80 \GeVc \jpt intervals. 

The EEC is unfolded in both \pPb and \pp events to evaluate the impact of the detector-effect correction method. In both datasets, the unfolded EEC generally shows $\leq 2 \%$ difference from bin-by-bin corrected results, and the difference is assigned as an \rl-dependent systematic uncertainty. 

In \pPb events, the uncertainty from the event-generator dependence is determined by comparing the correction factors extracted from events that are simulated with DPMJET~\cite{dpmjet} and PYTHIA+HIJING~\cite{pythia8, hijing} and propagated through a GEANT3 model of ALICE in both cases. The uncertainty thus obtained is $\leq4$\%. In \pp events, the event-generator dependence uncertainty is determined by comparing the correction factors extracted from events simulated with PYTHIA~\cite{pythia8} and Herwig~\cite{herwig}, with parameterized detector responses included, resulting in $\sim2$\% variation. An \rl-dependent uncertainty is used for both \pPb and \pp events. 

The systematic uncertainties are assumed to be independent from each other for a given collision system and are added in quadrature, resulting in a total uncertainty of $<$ 5\% for most \rl values, and never exceeding 10\% in \pp events and 15\% in \pPb events.

For the \pPb/\pp EEC ratio, the single-track efficiency and generator-dependence uncertainties from \pPb and \pp events are treated as uncorrelated, and added in quadrature. The remaining sources of uncertainty (pair efficiency, background subtraction, and correction method) are treated as correlated uncertainties and evaluated separately for the ratio. The total systematic uncertainty on the ratio is 2–4\% for 20–40 \GeVc jets except at small \rl, where it rises to about 10\%. In 40–60 \GeVc jets, it is similarly 2–4\% except at small \rl, where it rises to 10\%, driven by the pair-efficiency uncertainties. In 60–80 \GeVc jets, the total uncertainty is generally $\sim4$\%.

For the rapidity- and event–activity-dependent EECs in \pPb data, the same sources of systematic uncertainties are evaluated. They are all treated as correlated in the calculation of ratios of the EEC in different classes of \pPb data. For both these differential studies, the uncertainty is $\sim2$\% for 20–40 \GeVc jets except at small \rl, where it rises to about 10\%.

%% file: Ch5_Results.tex
\section{Results}

\begin{figure}[h!]
    \centering
    \includegraphics[width=\linewidth]{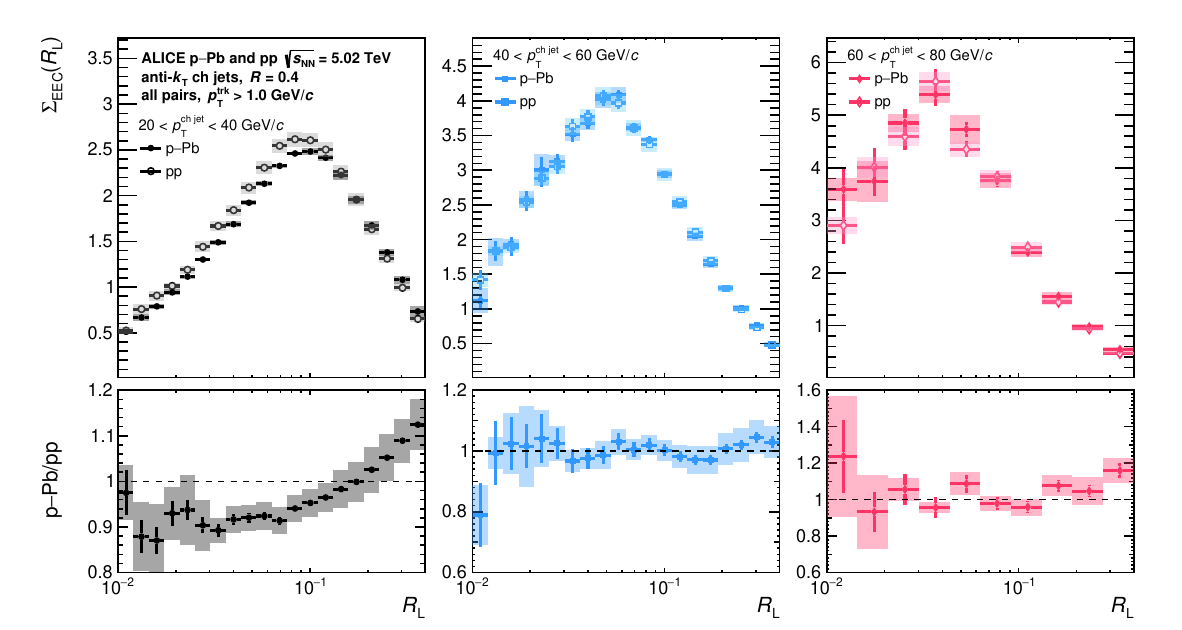}
    \caption{Top: The fully corrected \EEC in \pPb and \pp collisions as a function of \rl in 20–40, 40–60, and 60–80 \GeVc jets. Bottom: The ratio of \EEC in \pPb to \pp. Systematic uncertainties are shown with shaded boxes around the data points.}
    \label{fig:ratios}
\end{figure}

The EEC distributions for \pPb and \pp collisions in the \jpt ranges of 20–40, 40–60, and 60–80 \GeVc are shown in the top panels of Fig.~\ref{fig:ratios}. The same distinct \rl dependencies at the large-angle and small-angle limits are seen in \pPb as in \pp collisions. The large \rl region is driven by early-time jet evolution, characterized by the wide-angle splittings of the parton shower. The small \rl region is determined by late-time dynamics, dominated by the non-perturbative physics of colorless hadrons. The intermediate region features a striking peak, which has been suspected to be a signature of the confinement and hadronization process~\cite{opencmsdata}. Additionally, the same \jpt dependence is seen in \pPb as in \pp collisions. Namely, the perturbative region extends to smaller \rl in higher-\pt jets, leading to a shift in the EEC peak to lower \rl. 

Despite these general similarities, however, inspecting the \pPb/\pp EEC ratio in the bottom panels of Fig.~\ref{fig:ratios} reveals some differences. In the 20–40 \GeVc interval, the EEC shows an enhancement at large \rl, as well as a suppression at small \rl, relative to \pp events. Between $0.03 < \rl < 0.08$, the \pPb/\pp ratio is smaller than unity (where unity corresponds to the
null hypothesis) by 2.4–3.4$\sigma$. At the largest \rl value reported here, the \pPb/\pp ratio exceeds unity by 2.2$\sigma$. This shift towards larger angles is suggestive of a broadening of the energy flow, although alternative explanations—including changes in soft fragmentation relative to \pp collisions—cannot be excluded. This difference is not present in the two higher \jpt intervals. 

\subsection{Track \pt threshold dependence}
\label{sec:trackpt}
\begin{figure}[h!]
    \centering
    \includegraphics[width=\linewidth]{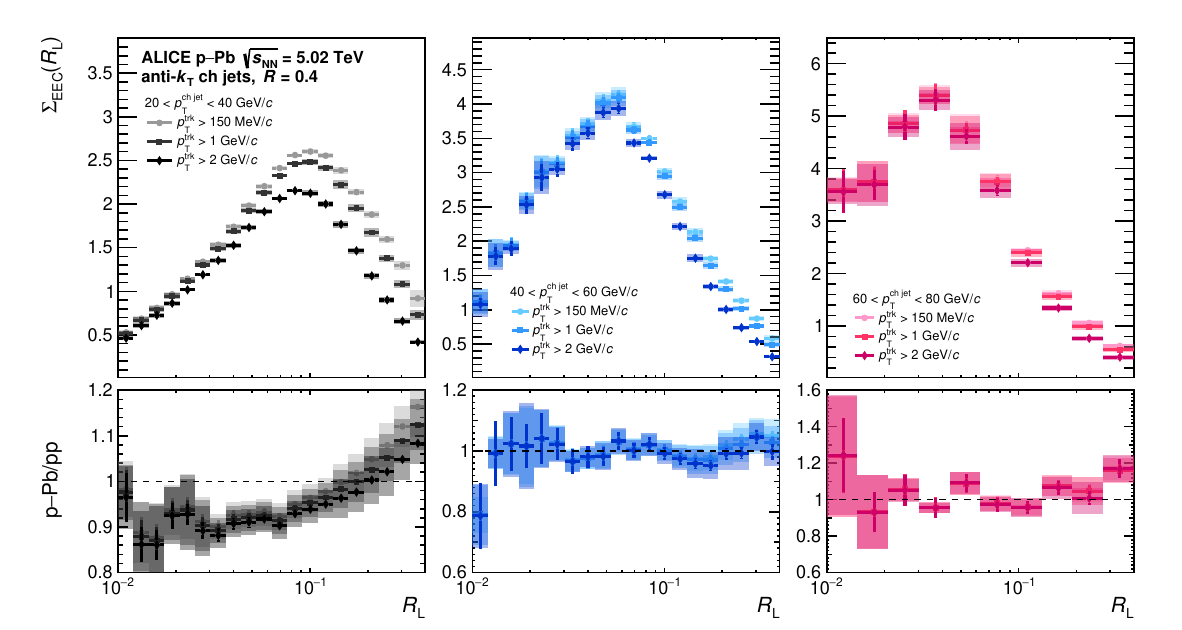}
    \caption{Top: The fully corrected \EEC in \pPb collisions as a function of \rl, with a varying \pttrk threshold (150 \MeVc, 1 \GeVc, and 2 \GeVc), in 20–40, 40–60, and 60–80 \GeVc jets. Bottom: The ratio of \EEC in \pPb to \pp with different \pttrk thresholds, where in the \pp reference the \pttrk threshold is also varied. Systematic uncertainties are shown with shaded boxes around the data points.}
    \label{fig:trkpt}
\end{figure}

As explained in section~\ref{jet-sel}, only constituents inside the reconstructed jets with \pttrk $>$ 1 \GeVc enter the EEC. This threshold can be varied to understand the impact of the jet's soft content on the EEC. Note that this does not change the jet \pt, as that value is always determined using tracks with \pt $\geq 150$ \MeVc. Figure~\ref{fig:trkpt} shows how the EEC (top panel) and the \pPb/\pp ratio (bottom panel) change after raising the \pttrk threshold to 2 \GeVc and decreasing it to 150 \MeVc. In the upper panels, it is clear that decreasing the \pttrk threshold increases the EEC level, and seems to shift the EEC peak to slightly higher \rl. This is most visible at lower jet \pt. In the \pPb/\pp ratio, varying the \pttrk value appears to tune the enhancement in 20–40 \GeVc jets, while leaving the small \rl behavior unchanged. Including softer tracks inside the jet increases the enhancement of the EEC ratio and shifts its onset to smaller \rl. However, the \pPb/\pp ratio is in agreement within error bars between all three \pttrk thresholds across the entire \rl range.

\subsection{Differential EEC studies}
\subsubsection{Detailed study of \jpt-dependent modification in 20–40 \GeVc jets}
\begin{figure}[h!]
    \centering
    \includegraphics[width=\linewidth]{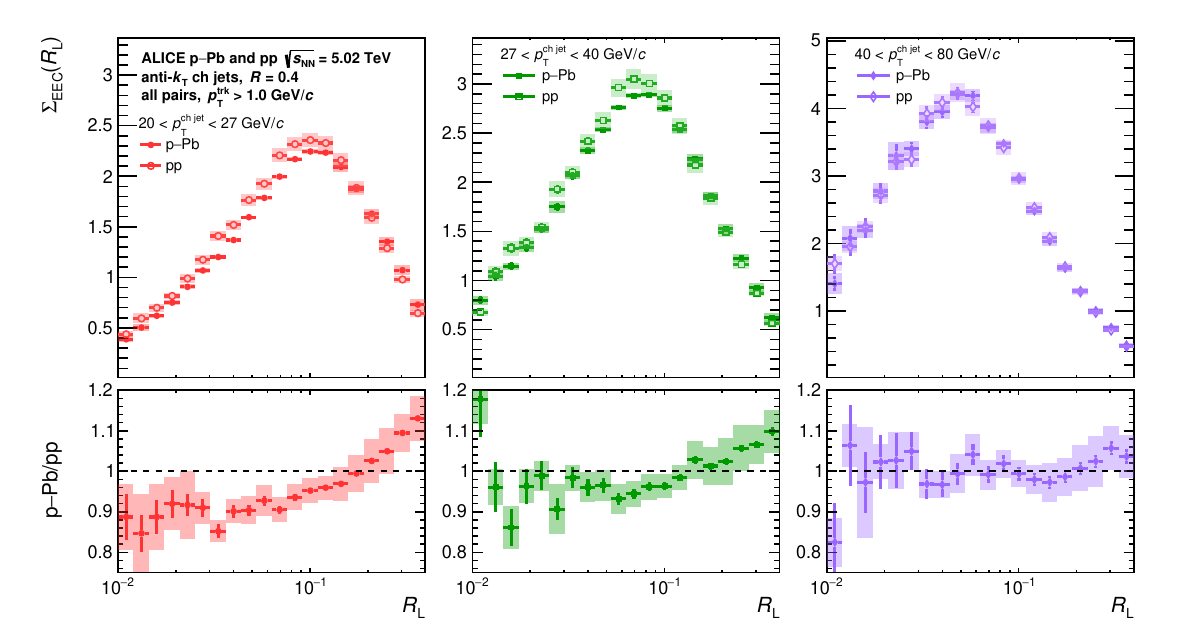}
    \caption{Top: The fully corrected \EEC in \pPb and \pp collisions as a function of \rl, in 20–27, 27–40, and 40–80 \GeVc jets. Bottom: The ratio of \EEC in \pPb to \pp. Systematic uncertainties are shown with shaded boxes around the data points.}
    \label{fig:splitbin}
\end{figure}

To understand more granularly the modification observed in 20–40 \GeVc jets, the EEC is measured in jets in 20–27 and 27–40 \GeVc intervals. This split was chosen in order to have roughly the same number of jets in each interval. The results are shown in the left two panels of Fig.~\ref{fig:splitbin}. The \pPb EEC differs from the \pp EEC in both \jpt intervals, but there is a clear \jpt dependence: the modification is stronger in 20–27 \GeVc jets than in 27–40 \GeVc jets. Additionally, Fig.~\ref{fig:splitbin} also shows the EEC in \pp and \pPb collisions for a combined 40–80 \GeVc \jpt interval. In this combined range, there is no evidence for a modification of the EEC, which is consistent with results in the individual ranges.

\subsubsection{Dependence on jet pseudorapidity}
Next, the dependence of EEC on pseudorapidity is considered in the 20$<$\jpt$<$40 \GeVc interval. The overall ${\rm d}N/{\rm d}\eta$ distribution in \pPb collisions is asymmetric due to the asymmetric collision system~\cite{dndeta-ppb}. The pseudorapidity density skews higher in the Pb-going direction (i.e. in the direction the Pb ion beam travels) and lower in the p-going direction (i.e. the direction the proton beam). To probe for local particle density effects, the EEC is constructed for Pb-going jets ($0<\eta_{\rm jet}<0.5$) and p-going jets ($-0.5<\eta_{\rm jet}<0$). The \EEC$_{\rm p-going}$/\EEC$_{\rm Pb-going}$ ratio is shown in the upper panel of Fig.~\ref{fig:diffratios}. The p-going and Pb-going EECs agree within 5\%, showing no evidence of pseudorapidity dependence. It appears that the ${\rm d}N/{\rm d}\eta$ asymmetry does not affect the EEC. This also indicates that the handling of EEC background is robust, because the overall increase in ${\rm d}N/{\rm d}\eta$ in the Pb-going direction does not manifest as a change in the EEC.

\subsubsection{Dependence on forward multiplicity}
\begin{figure}[h!]
    \centering
    \includegraphics[width=0.6\linewidth]{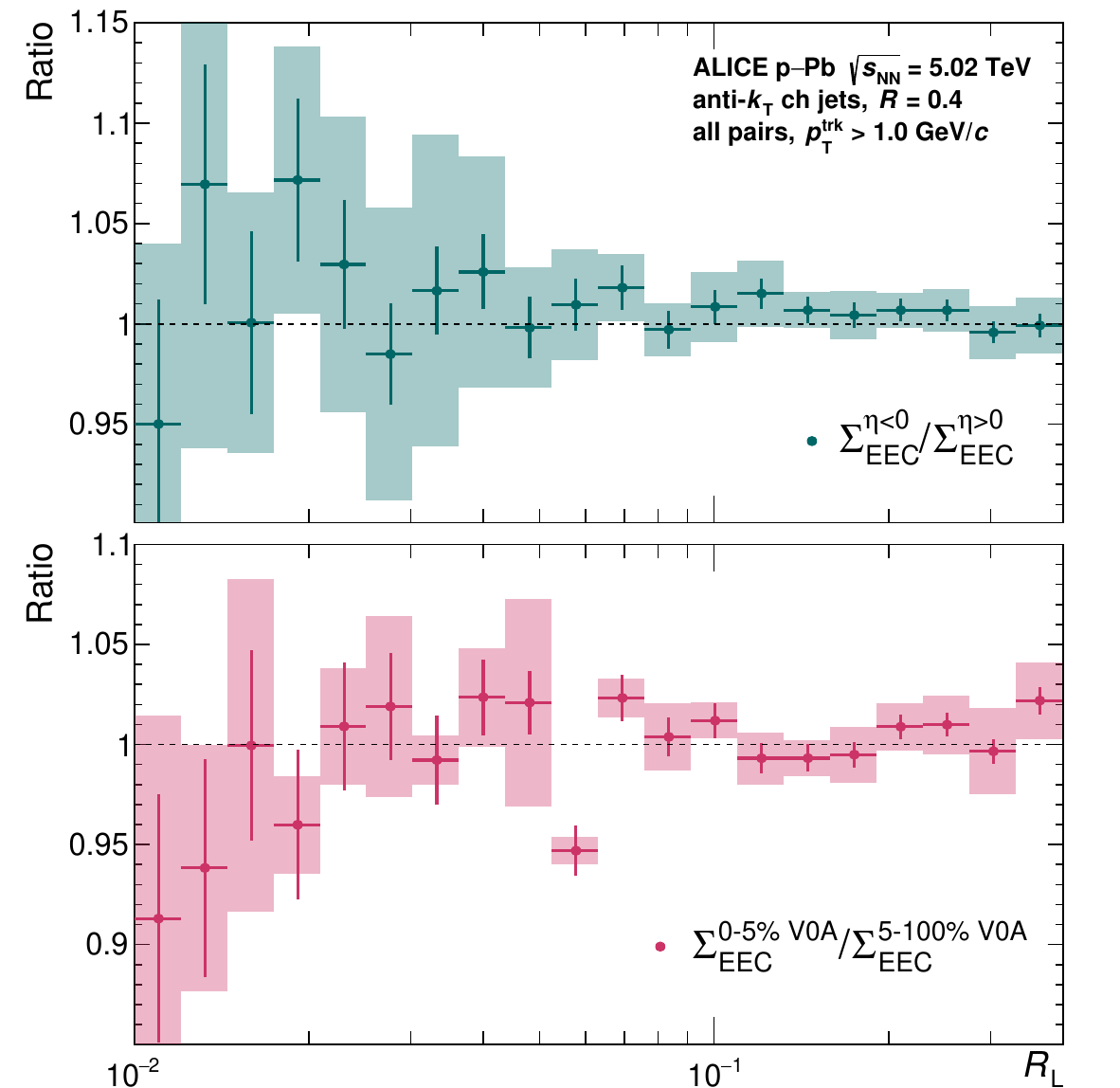}
    \caption{Top: The ratio of p-going ($\eta_{\rm jet}<0$) to Pb-going ($\eta_{\rm jet}>0$) \EEC in 20–40 \GeVc \pPb jets. Bottom: The ratio of \EEC in high-multiplicity events (0–5\% V0A) to \EEC in low-multiplicity events (5–100\% V0A), in 20–40 \GeVc \pPb jets.}
    \label{fig:diffratios}
\end{figure}

The dependence of the modification on event activity is tested by measuring the EEC in two event-activity classes, determined using forward charged-particle multiplicity measured with the V0A detector. The V0A amplitude is frequently used as a centrality estimator in \pPb measurements~\cite{centrality-ppb}. Here, it is used to define two percentile-based multiplicity classes, high-multiplicity (HM) and low-multiplicity (LM). The low-multiplicity events are defined as all those below the 95\% percentile (5–100\% V0A), and high-multiplicity events as those with the 0–5\% highest V0A activity. The EEC is separately constructed for jets in each event class, and the event-activity dependence is studied by taking the \EEC$_{\rm HM}$/\EEC$_{\rm LM}$ ratio, shown in the bottom panel of Fig.~\ref{fig:diffratios}. Here, as well, the EECs agree within 5\% for most of the reported \rl range, indicating no evidence of event-activity dependence. This suggests that the initial collision geometry does not influence the EEC in \pPb events.

\subsection{Studying the transition region}

In \pp collisions, it was found that the EEC  aligns with pQCD calculations in the large \rl region, and with free hadron scaling in the small \rl region~\cite{alice-pp-eec}. The intermediate region, where the behavior shifts from mostly perturbative to mostly non-perturbative, has been referred to as a `transition region'.

\begin{figure}
    \centering
    \includegraphics[width=0.6\linewidth]{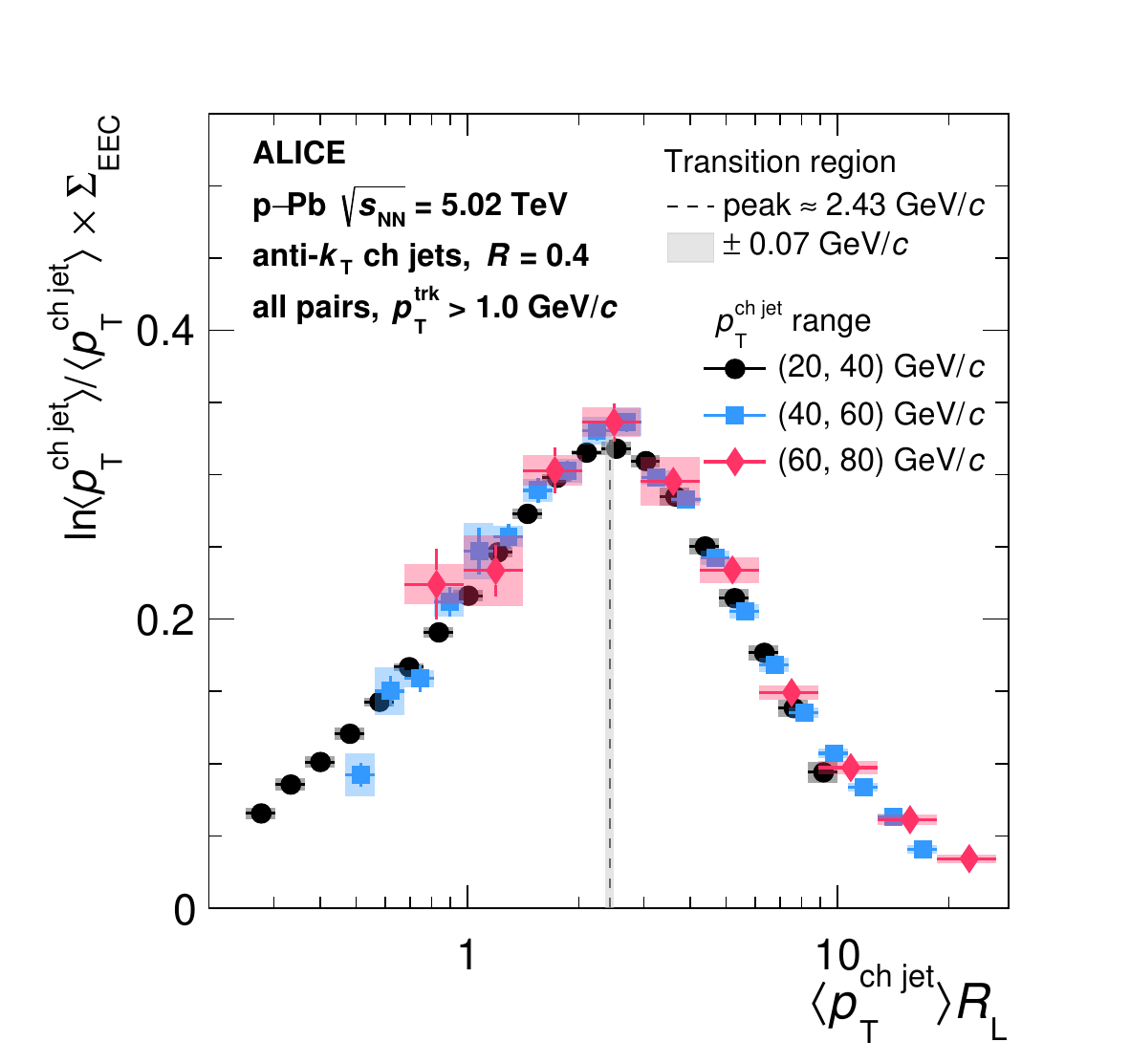}
    \caption{The normalized \EEC as a function of \ptrl. The logarithm $\ln\langle p_{\rm T}^{\rm ch\> jet}\rangle$ represents $\ln(\langle p_{\rm T}^{\rm ch\> jet}\rangle/({\rm GeV}/c))$. The dashed grey line corresponds to the maximum location of the distribution and the grey band corresponds to a $\pm0.07$ \GeVc uncertainty along the x-axis. }
    \label{fig:univ}
\end{figure}

As described in Ref.~\cite{alice-pp-eec}, the peak position ($R_{\rm L}^{\rm peak}$) and peak height ($\Sigma_{\rm EEC}^{\rm peak}$) were extracted for the measured \jpt intervals to study the dependence of the transition region on the jet energy in \pp collisions. It was found that ${\rl}^{\rm peak}$ was proportional to $1/\langle\jpt\rangle$, which is what one would expect if the \EEC peak corresponded to a fixed hadronization scale. Upon inspecting the EEC as a function of \ptrl, and scaling \EEC by $\ln\langle\jpt\rangle/\langle\jpt\rangle$, it was found that the \pp EEC curves collapsed into a common shape~\cite{alice-pp-eec}. To see if the same \jpt dependence of the peak position and height holds in \pPb collisions, these scalings are performed using the same methods described in detail in Ref.~\cite{alice-pp-eec}. The results are shown in Fig.~\ref{fig:univ}. From this plot, it is clear that $\ptrl^{\rm peak}$ is consistent between the three \jpt intervals, and has a value of 2.43$\pm$0.07 \GeVc in \pPb collisions, where the uncertainty includes both statistical and systematic contributions propagated through the peak-extraction fit. In the previously published \pp EEC measurement from ALICE (without underlying event subtraction), the peak position was reported to be 2.40$\pm$0.10 \GeVc~\cite{alice-pp-eec}. Thus a consistent peak position is found not only across jet \pt intervals, but also between \pp and \pPb collisions. The peak height $\ln\langle\jpt\rangle/\langle\jpt\rangle\Sigma_{\rm EEC}^{\rm peak}$ is found to be 0.33$\pm$0.01, which is quite close to the value of 0.31$\pm$0.01 found in \pp~\cite{alice-pp-eec}. Here as well, the uncertainty includes statistical and systematic contributions.

\subsection{Model comparisons}
\begin{figure}[h!]
    \centering
    \includegraphics[width=0.7\linewidth]{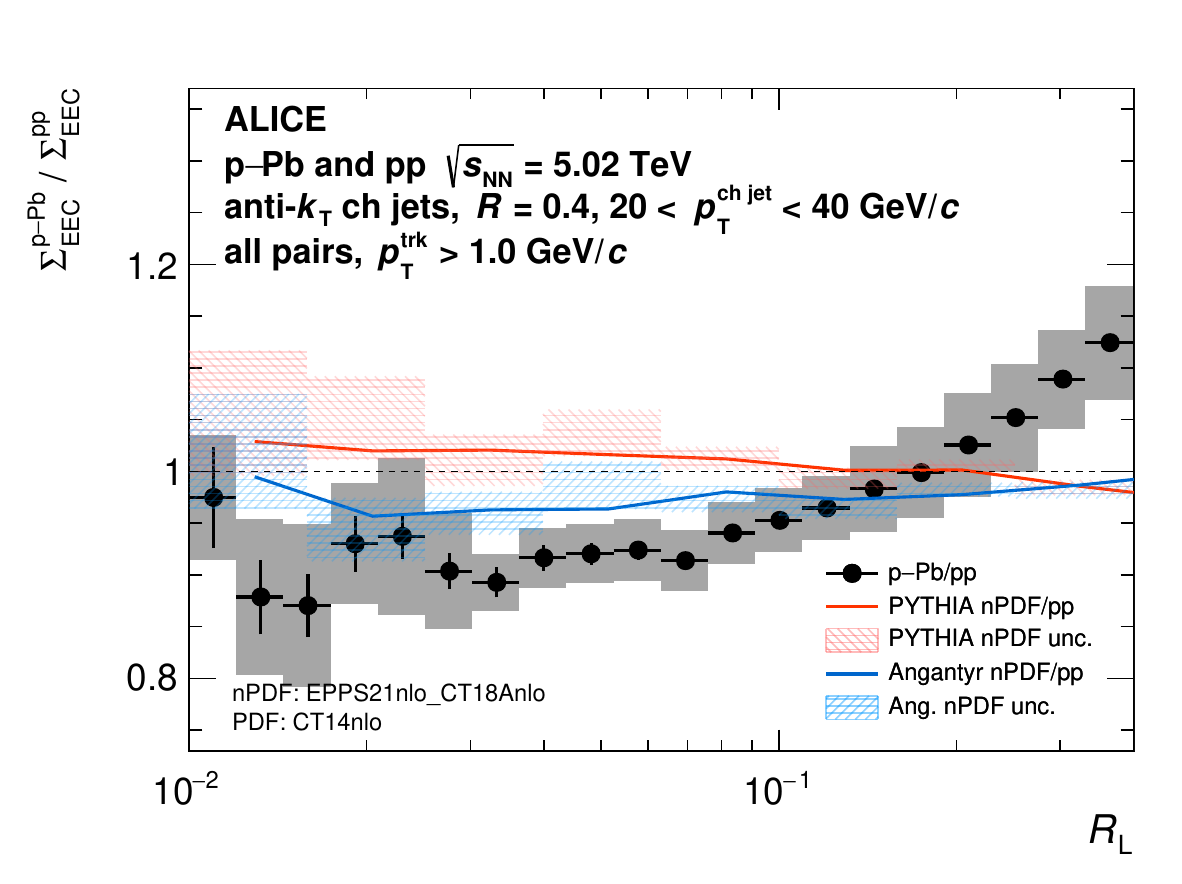}
    \caption{The ratio of \pPb to \pp \EEC in 20–40 \GeVc jets, with comparison to PYTHIA with nPDF (red)~\cite{epps21} and PYTHIA Angantyr (blue)~\cite{angantyr}.}
    \label{fig:npdf}
\end{figure}

To study the possible origins of the \pPb modification, the data are compared to two Monte Carlo event generators in Fig.~\ref{fig:npdf}. First, to isolate the effect of the nuclear parton distribution function (nPDF) alone on the EEC, PYTHIA8~\cite{pythia8} is used with the LHAPDF6 package~\cite{lhapdf6}, which supplies nPDFs to describe the non-perturbative initial state. In particular, the CT14 NLO free-proton PDF set is used for one of the colliding protons~\cite{ct14-nlo-pdf}, and the EPPS21 NLO nPDF set is used for the other proton~\cite{epps21}. For the \pp reference, the CT14 NLO PDF set is used for both colliding protons. The EEC is not modified by the introduction of the nPDF, within the uncertainties of the nPDF set. 

The data are also compared to PYTHIA Angantyr~\cite{angantyr}, which more comprehensively models multi-parton interactions and diffractive excitation because of fluctuations in nucleon partonic substructure. Angantyr provides a more complete simulation of \pPb collisions, and it predicts a modification to the EEC. The \pPb/\pp EEC ratio in Angantyr indicates a suppression at small \rl, but does not capture the enhancement at large \rl measured in the data. The initial-state differences introduced into \pPb collisions by these models are insufficient to explain the measured EEC modification relative to \pp collisions, suggesting a potential role for final-state effects when jets propagate through CNM.

\begin{figure}[h!]
    \centering
    \includegraphics[width=0.7\linewidth]{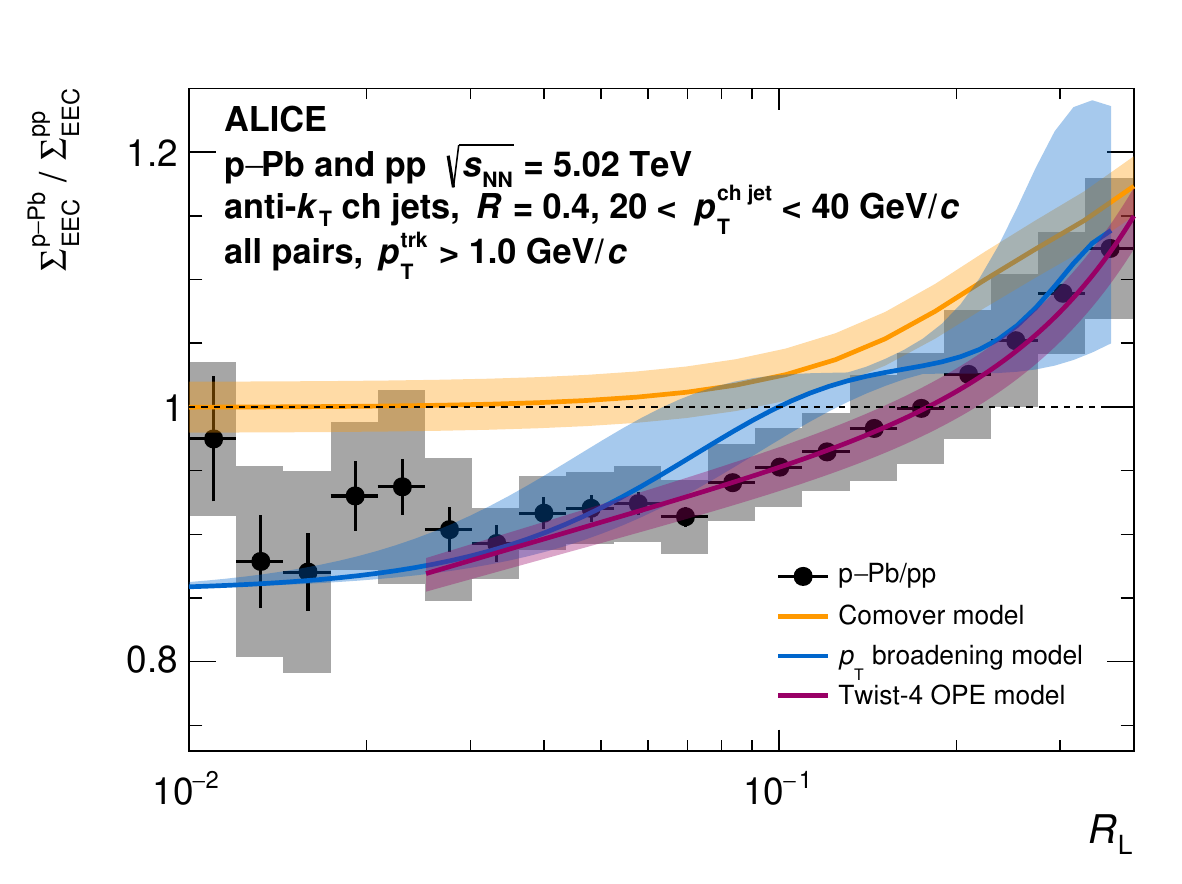}
    \caption{The ratio of \pPb to \pp \EEC in 20–40 \GeVc jets, with comparison to theoretical models with comovers~\cite{berndt-eec}, \pt broadening~\cite{zhongbo-eec}, and twist-4 OPE corrections~\cite{andres2024-pA}.}
    \label{fig:theory}
\end{figure}

Figure~\ref{fig:theory} shows the ratio of the EEC in \pPb to \pp, compared to three different theoretical models incorporating final-state effects. The models implement partonic comovers~\cite{berndt-eec}, transverse-momentum broadening~\cite{zhongbo-eec}, and twist-4 OPE corrections~\cite{andres2024-pA}. 

The comover model uses a higher-twist formalism to account for interactions between the jet and comovers, which are particles produced in interactions between the remnants of the proton and the nucleus in a \pA collision. This model does not attempt to capture the non-perturbative region of the EEC, so it misses the suppression in the ratio at small \rl. Nevertheless it does qualitatively describe the trends in the data at large \rl. 

The \pt-broadening model uses transverse-momentum-dependent (TMD) factorization to describe the non-perturbative \rl region, and a leading-logarithmic calculation of a hard function and a jet function to capture the perturbative region. To model \pPb collisions, the perturbative calculation includes an nPDF and a transport parameter~\cite{quenching-cnm}, and the non-perturbative sector accounts for multiple scattering and the resulting \pt broadening. This model does reasonably well with the 20–40 \GeVc data, capturing small \rl suppression as well as large \rl enhancement. The enhancement is largely driven by the perturbative modification. However, this model also predicts a modification in the 60–80 \GeVc jet \pt interval, which is not observed in the data presented earlier. 

The third model exploits the fact that energy correlators can be expressed in terms of a light-ray operator product expansion (OPE), which turns them into a series in \rl (called $\theta$ in Ref.~\cite{andres2024-pA}). The perturbative scaling of the EEC in \pp collisions is related to the anomalous dimensions of twist-2 light ray operators. In \pA events, the modification of the EEC is related to the matrix elements of twist-4 light-ray operators. In this model, the twist-2 and twist-4 terms
are fit to the 20–40 \GeVc data, leading to excellent agreement in this jet \pt interval. However, the authors also find good agreement at higher jet \pt, validating the \pt-dependence of this model.

%% file: Ch6_Conclusions.tex
\section{Conclusions}

This paper reports the first measurement of the two-point energy–energy correlator inside $R=0.4$ anti-$k_{\mathrm{T}}$ charged-particle jets in \pPb collisions at \snn = 5.02 TeV. New background-subtracted measurements of the EEC in \pp collisions at \s = 5.02 TeV are also shown. For jets with $p_{\mathrm{T}}$ between 40–80 \GeVc, the EEC in \pPb collisions agrees with the \pp baseline within uncertainties. In the 20–40 \GeVc interval, however, an enhancement is observed at large $R_{\mathrm{L}}$ and a suppression at small \rl, indicating a modification of the angular distribution of energy flow.

This modification is still visible even if the track \pt threshold for EEC pairs is increased to 2 \GeVc. The modification shows no significant dependence on jet pseudorapidity or forward multiplicity, suggesting that it is neither related to local particle density nor collision geometry effects. The modification is also not reproduced by PYTHIA simulations with nuclear PDFs or by PYTHIA Angantyr. These comparisons suggest that initial-state effects alone are insufficient to explain the observed pattern, indicating a possible role for final-state interactions in \pPb jets.

Theoretical frameworks have been developed to propose possible origins. Current models incorporating final-state effects qualitatively capture measured features of the data, but do not yet provide a complete quantitative description. Additional differential measurements and refined theoretical calculations will be essential to disentangle CNM-induced broadening from modifications of soft fragmentation in small systems.

%% file: fa_2026-06-03_Opt_C.tex

The ALICE Collaboration would like to thank all its engineers and technicians for their invaluable contributions to the construction of the experiment and the CERN accelerator teams for the outstanding performance of the LHC complex.
The ALICE Collaboration gratefully acknowledges the resources and support provided by all Grid centres and the Worldwide LHC Computing Grid (WLCG) collaboration.
The ALICE Collaboration acknowledges the following funding agencies for their support in building and running the ALICE detector:
A. I. Alikhanyan National Science Laboratory (Yerevan Physics Institute) Foundation (ANSL), State Committee of Science and World Federation of Scientists (WFS), Armenia;
Austrian Academy of Sciences, Austrian Science Fund (FWF): [M 2467-N36] and Nationalstiftung f\"{u}r Forschung, Technologie und Entwicklung, Austria;
Ministry of Communications and High Technologies, National Nuclear Research Center, Azerbaijan;
Rede Nacional de Física de Altas Energias (Renafae), Financiadora de Estudos e Projetos (Finep), Funda\c{c}\~{a}o de Amparo \`{a} Pesquisa do Estado de S\~{a}o Paulo (FAPESP) and The Sao Paulo Research Foundation  (FAPESP), Brazil;
Bulgarian Ministry of Education and Science, within the National Roadmap for Research Infrastructures 2020-2027 (object CERN), Bulgaria;
Ministry of Education of China (MOEC) , Ministry of Science \& Technology of China (MSTC) and National Natural Science Foundation of China (NSFC), China;
Ministry of Science and Education and Croatian Science Foundation, Croatia;
Centro de Aplicaciones Tecnol\'{o}gicas y Desarrollo Nuclear (CEADEN), Cubaenerg\'{\i}a, Cuba;
Ministry of Education, Youth and Sports of the Czech Republic, Czech Republic;
The Danish Council for Independent Research | Natural Sciences, the VILLUM FONDEN and Danish National Research Foundation (DNRF), Denmark;
Helsinki Institute of Physics (HIP), Finland;
Commissariat \`{a} l'Energie Atomique (CEA) and Institut National de Physique Nucl\'{e}aire et de Physique des Particules (IN2P3) and Centre National de la Recherche Scientifique (CNRS), France;
Bundesministerium f\"{u}r Forschung, Technologie und Raumfahrt (BMFTR) and GSI Helmholtzzentrum f\"{u}r Schwerionenforschung GmbH, Germany;
National Research, Development and Innovation Office, Hungary;
Department of Atomic Energy Government of India (DAE), Department of Science and Technology, Government of India (DST), University Grants Commission, Government of India (UGC) and Council of Scientific and Industrial Research (CSIR), India;
National Research and Innovation Agency - BRIN, Indonesia;
Istituto Nazionale di Fisica Nucleare (INFN), Italy;
Japanese Ministry of Education, Culture, Sports, Science and Technology (MEXT) and Japan Society for the Promotion of Science (JSPS) KAKENHI, Japan;
Consejo Nacional de Ciencia (CONACYT) y Tecnolog\'{i}a, through Fondo de Cooperaci\'{o}n Internacional en Ciencia y Tecnolog\'{i}a (FONCICYT) and Direcci\'{o}n General de Asuntos del Personal Academico (DGAPA), Mexico;
Nederlandse Organisatie voor Wetenschappelijk Onderzoek (NWO), Netherlands;
The Research Council of Norway, Norway;
Pontificia Universidad Cat\'{o}lica del Per\'{u}, Peru;
Ministry of Science and Higher Education, National Science Centre and WUT ID-UB, Poland;
Korea Institute of Science and Technology Information and National Research Foundation of Korea (NRF), Republic of Korea;
Ministry of Education and Scientific Research, Institute of Atomic Physics, Ministry of Research and Innovation and Institute of Atomic Physics and Universitatea Nationala de Stiinta si Tehnologie Politehnica Bucuresti, Romania;
Ministerstvo skolstva, vyskumu, vyvoja a mladeze SR, Slovakia;
National Research Foundation of South Africa, South Africa;
Swedish Research Council (VR) and Knut \& Alice Wallenberg Foundation (KAW), Sweden;
European Organization for Nuclear Research, Switzerland;
Suranaree University of Technology (SUT), National Science and Technology Development Agency (NSTDA) and National Science, Research and Innovation Fund (NSRF via PMU-B B05F650021), Thailand;
Turkish Energy, Nuclear and Mineral Research Agency (TENMAK), Turkey;
National Academy of  Sciences of Ukraine, Ukraine;
Science and Technology Facilities Council (STFC), United Kingdom;
National Science Foundation of the United States of America (NSF) and United States Department of Energy, Office of Nuclear Physics (DOE NP), United States of America.
In addition, individual groups or members have received support from:
Czech Science Foundation (grant no. 23-07499S), Czech Republic;
FORTE project, reg.\ no.\ CZ.02.01.01/00/22\_008/0004632, Czech Republic, co-funded by the European Union, Czech Republic;
European Research Council (grant no. 101220549), European Union;
Deutsche Forschungs Gemeinschaft (DFG, German Research Foundation) ``Neutrinos and Dark Matter in Astro- and Particle Physics'' (grant no. SFB 1258), Germany;
CONVECS project, CUP C97H23001700002 FESR 2021-2027 program, Italy.

%% file: Alice_Authorlist_2026-06-03_Opt_C.tex
\begin{flushleft} 
\small

D.A.H.~Abdallah\,\orcidlink{0000-0003-4768-2718}\,$^{\rm 134}$, 
I.J.~Abualrob\,\orcidlink{0009-0005-3519-5631}\,$^{\rm 112}$, 
S.~Acharya\,\orcidlink{0000-0002-9213-5329}\,$^{\rm 49}$, 
K.~Agarwal\,\orcidlink{0000-0001-5781-3393}\,$^{\rm II,}$$^{\rm 23}$, 
G.~Aglieri Rinella\,\orcidlink{0000-0002-9611-3696}\,$^{\rm 32}$, 
L.~Aglietta\,\orcidlink{0009-0003-0763-6802}\,$^{\rm 24}$, 
N.~Agrawal\,\orcidlink{0000-0003-0348-9836}\,$^{\rm 25}$, 
Z.~Ahammed\,\orcidlink{0000-0001-5241-7412}\,$^{\rm 132}$, 
S.~Ahmad\,\orcidlink{0000-0003-0497-5705}\,$^{\rm 15}$, 
Z.~Akbar$^{\rm 79}$
% ...

\section*{Affiliation Notes}

$^{\rm I}$ Deceased\\
$^{\rm II}$ Also at: INFN Trieste\\
$^{\rm III}$ Also at: Fondazione Bruno Kessler (FBK), Trento, Italy\\
$^{\rm IV}$ Also at: Czech Technical University in Prague (CZ)\\
$^{\rm V}$ Also at: Instituto de Fisica da Universidade de Sao Paulo\\
$^{\rm VI}$ Also at: Dipartimento DET del Politecnico di Torino, Turin, Italy\\
$^{\rm VII}$ Also at: Department of Applied Physics, Aligarh Muslim University, Aligarh, India\\
$^{\rm VIII}$ Also at: Institute of Theoretical Physics, University of Wroclaw, Poland\\
$^{\rm IX}$ Also at: Facultad de Ciencias, Universidad Nacional Aut\'{o}noma de M\'{e}xico, Mexico City, Mexico\\

\section*{Collaboration Institutes}

$^{1}$ A.I. Alikhanyan National Science Laboratory (Yerevan Physics Institute) Foundation, Yerevan, Armenia\\
$^{2}$ AGH University of Krakow, Cracow, Poland\\
$^{3}$ Bogolyubov Institute for Theoretical Physics, National Academy of Sciences of Ukraine, Kyiv, Ukraine\\
$^{4}$ Bose Institute, Department of Physics  and Centre for Astroparticle Physics and Space Science (CAPSS), Kolkata, India\\
$^{5}$ California Polytechnic State University, San Luis Obispo, California, United States\\
$^{6}$ Central China Normal University, Wuhan, China\\
$^{7}$ Centro de Aplicaciones Tecnol\'{o}gicas y Desarrollo Nuclear (CEADEN), Havana, Cuba\\
$^{8}$ Centro de Investigaci\'{o}n y de Estudios Avanzados (CINVESTAV), Mexico City and M\'{e}rida, Mexico\\
$^{9}$ Chicago State University, Chicago, Illinois, United States\\
$^{10}$ China Nuclear Data Center, China Institute of Atomic Energy, Beijing, China\\
$^{11}$ China University of Geosciences, Wuhan, China\\
$^{12}$ Chungbuk National University, Cheongju, Republic of Korea\\
$^{13}$ Comenius University Bratislava, Faculty of Mathematics, Physics and Informatics, Bratislava, Slovak Republic\\
$^{14}$ Creighton University, Omaha, Nebraska, United States\\
$^{15}$ Department of Physics, Aligarh Muslim University, Aligarh, India\\
$^{16}$ Department of Physics, Pusan National University, Pusan, Republic of Korea\\
$^{17}$ Department of Physics, Sejong University, Seoul, Republic of Korea\\
$^{18}$ Department of Physics, University of California, Berkeley, California, United States\\
$^{19}$ Department of Physics, University of Oslo, Oslo, Norway\\
$^{20}$ Department of Physics and Technology, University of Bergen, Bergen, Norway\\
$^{21}$ Dipartimento di Fisica, Universit\`{a} di Pavia, Pavia, Italy\\
$^{22}$ Dipartimento di Fisica dell'Universit\`{a} and Sezione INFN, Cagliari, Italy\\
$^{23}$ Dipartimento di Fisica dell'Universit\`{a} and Sezione INFN, Trieste, Italy\\
$^{24}$ Dipartimento di Fisica dell'Universit\`{a} and Sezione INFN, Turin, Italy\\
$^{25}$ Dipartimento di Fisica e Astronomia dell'Universit\`{a} and Sezione INFN, Bologna, Italy\\
$^{26}$ Dipartimento di Fisica e Astronomia dell'Universit\`{a} and Sezione INFN, Catania, Italy\\
$^{27}$ Dipartimento di Fisica e Astronomia dell'Universit\`{a} and Sezione INFN, Padova, Italy\\
$^{28}$ Dipartimento di Fisica `E.R.~Caianiello' dell'Universit\`{a} and Gruppo Collegato INFN, Salerno, Italy\\
$^{29}$ Dipartimento DISAT del Politecnico and Sezione INFN, Turin, Italy\\
$^{30}$ Dipartimento di Scienze MIFT, Universit\`{a} di Messina, Messina, Italy\\
$^{31}$ Dipartimento Interateneo di Fisica `M.~Merlin' and Sezione INFN, Bari, Italy\\
$^{32}$ European Organization for Nuclear Research (CERN), Geneva, Switzerland\\
$^{33}$ Faculty of Electrical Engineering, Mechanical Engineering and Naval Architecture, University of Split, Split, Croatia\\
$^{34}$ Faculty of Nuclear Sciences and Physical Engineering, Czech Technical University in Prague, Prague, Czech Republic\\
$^{35}$ Faculty of Physics, Sofia University, Sofia, Bulgaria\\
$^{36}$ Faculty of Science, P.J.~\v{S}af\'{a}rik University, Ko\v{s}ice, Slovak Republic\\
$^{37}$ Faculty of Technology, Environmental and Social Sciences, Bergen, Norway\\
$^{38}$ Frankfurt Institute for Advanced Studies, Johann Wolfgang Goethe-Universit\"{a}t Frankfurt, Frankfurt, Germany\\
$^{39}$ Fudan University, Shanghai, China\\
$^{40}$ Gauhati University, Department of Physics, Guwahati, India\\
$^{41}$ Helmholtz-Institut f\"{u}r Strahlen- und Kernphysik, Rheinische Friedrich-Wilhelms-Universit\"{a}t Bonn, Bonn, Germany\\
$^{42}$ Helsinki Institute of Physics (HIP), Helsinki, Finland\\
$^{43}$ High Energy Physics Group,  Universidad Aut\'{o}noma de Puebla, Puebla, Mexico\\
$^{44}$ Horia Hulubei National Institute of Physics and Nuclear Engineering, Bucharest, Romania\\
$^{45}$ HUN-REN Wigner Research Centre for Physics, Budapest, Hungary\\
$^{46}$ Indian Institute of Technology Bombay (IIT), Mumbai, India\\
$^{47}$ Indian Institute of Technology Indore, Indore, India\\
$^{48}$ INFN, Laboratori Nazionali di Frascati, Frascati, Italy\\
$^{49}$ INFN, Sezione di Bari, Bari, Italy\\
$^{50}$ INFN, Sezione di Bologna, Bologna, Italy\\
$^{51}$ INFN, Sezione di Cagliari, Cagliari, Italy\\
$^{52}$ INFN, Sezione di Catania, Catania, Italy\\
$^{53}$ INFN, Sezione di Padova, Padova, Italy\\
$^{54}$ INFN, Sezione di Pavia, Pavia, Italy\\
$^{55}$ INFN, Sezione di Torino, Turin, Italy\\
$^{56}$ INFN, Sezione di Trieste, Trieste, Italy\\
$^{57}$ Inha University, Incheon, Republic of Korea\\
$^{58}$ Institute for Gravitational and Subatomic Physics (GRASP), Utrecht University/Nikhef, Utrecht, Netherlands\\
$^{59}$ Institute of Experimental Physics, Slovak Academy of Sciences, Ko\v{s}ice, Slovak Republic\\
$^{60}$ Institute of Physics, Homi Bhabha National Institute, Bhubaneswar, India\\
$^{61}$ Institute of Physics of the Czech Academy of Sciences, Prague, Czech Republic\\
$^{62}$ Institute of Space Science (ISS), Bucharest, Romania\\
$^{63}$ Institut f\"{u}r Kernphysik, Johann Wolfgang Goethe-Universit\"{a}t Frankfurt, Frankfurt, Germany\\
$^{64}$ Instituto de Ciencias Nucleares, Universidad Nacional Aut\'{o}noma de M\'{e}xico, Mexico City, Mexico\\
$^{65}$ Instituto de F\'{i}sica, Universidade Federal do Rio Grande do Sul (UFRGS), Porto Alegre, Brazil\\
$^{66}$ Instituto de F\'{\i}sica, Universidad Nacional Aut\'{o}noma de M\'{e}xico, Mexico City, Mexico\\
$^{67}$ iThemba LABS, National Research Foundation, Somerset West, South Africa\\
$^{68}$ Jeonbuk National University, Jeonju, Republic of Korea\\
$^{69}$ Korea Institute of Science and Technology Information, Daejeon, Republic of Korea\\
$^{70}$ Laboratoire de Physique Subatomique et de Cosmologie, Universit\'{e} Grenoble-Alpes, CNRS-IN2P3, Grenoble, France\\
$^{71}$ Lawrence Berkeley National Laboratory, Berkeley, California, United States\\
$^{72}$ Lund University Department of Physics, Division of Particle Physics, Lund, Sweden\\
$^{73}$ Marietta Blau Institute, Vienna, Austria\\
$^{74}$ Nagasaki Institute of Applied Science, Nagasaki, Japan\\
$^{75}$ Nara Women{'}s University (NWU), Nara, Japan\\
$^{76}$ National Centre for Nuclear Research, Warsaw, Poland\\
$^{77}$ National Institute of Science Education and Research, Homi Bhabha National Institute, Jatni, India\\
$^{78}$ National Nuclear Research Center, Baku, Azerbaijan\\
$^{79}$ National Research and Innovation Agency - BRIN, Jakarta, Indonesia\\
$^{80}$ Niels Bohr Institute, University of Copenhagen, Copenhagen, Denmark\\
$^{81}$ Nikhef, National institute for subatomic physics, Amsterdam, Netherlands\\
$^{82}$ Nuclear Physics Group, STFC Daresbury Laboratory, Daresbury, United Kingdom\\
$^{83}$ Nuclear Physics Institute of the Czech Academy of Sciences, Husinec-\v{R}e\v{z}, Czech Republic\\
$^{84}$ Oak Ridge National Laboratory, Oak Ridge, Tennessee, United States\\
$^{85}$ Ohio State University, Columbus, Ohio, United States\\
$^{86}$ Physics department, Faculty of science, University of Zagreb, Zagreb, Croatia\\
$^{87}$ Physics Department, Panjab University, Chandigarh, India\\
$^{88}$ Physics Department, University of Jammu, Jammu, India\\
$^{89}$ Physics Program and International Institute for Sustainability with Knotted Chiral Meta Matter (WPI-SKCM$^{2}$), Hiroshima University, Hiroshima, Japan\\
$^{90}$ Physikalisches Institut, Eberhard-Karls-Universit\"{a}t T\"{u}bingen, T\"{u}bingen, Germany\\
$^{91}$ Physikalisches Institut, Ruprecht-Karls-Universit\"{a}t Heidelberg, Heidelberg, Germany\\
$^{92}$ Physik Department, Technische Universit\"{a}t M\"{u}nchen, Munich, Germany\\
$^{93}$ Politecnico di Bari and Sezione INFN, Bari, Italy\\
$^{94}$ Research Division and ExtreMe Matter Institute EMMI, GSI Helmholtzzentrum f\"ur Schwerionenforschung GmbH, Darmstadt, Germany\\
$^{95}$ Saga University, Saga, Japan\\
$^{96}$ Saha Institute of Nuclear Physics, Homi Bhabha National Institute, Kolkata, India\\
$^{97}$ School of Physics and Astronomy, University of Birmingham, Birmingham, United Kingdom\\
$^{98}$ Secci\'{o}n F\'{\i}sica, Departamento de Ciencias, Pontificia Universidad Cat\'{o}lica del Per\'{u}, Lima, Peru\\
$^{99}$ SUBATECH, IMT Atlantique, Nantes Universit\'{e}, CNRS-IN2P3, Nantes, France\\
$^{100}$ Sungkyunkwan University, Suwon City, Republic of Korea\\
$^{101}$ Suranaree University of Technology, Nakhon Ratchasima, Thailand\\
$^{102}$ Technical University of Ko\v{s}ice, Ko\v{s}ice, Slovak Republic\\
$^{103}$ The Henryk Niewodniczanski Institute of Nuclear Physics, Polish Academy of Sciences, Cracow, Poland\\
$^{104}$ The University of Texas at Austin, Austin, Texas, United States\\
$^{105}$ Universidad Aut\'{o}noma de Sinaloa, Culiac\'{a}n, Mexico\\
$^{106}$ Universidade de S\~{a}o Paulo (USP), S\~{a}o Paulo, Brazil\\
$^{107}$ Universidade Estadual de Campinas (UNICAMP), Campinas, Brazil\\
$^{108}$ Universidade Federal do ABC, Santo Andre, Brazil\\
$^{109}$ Universitatea Nationala de Stiinta si Tehnologie Politehnica Bucuresti, Bucharest, Romania\\
$^{110}$ University of Cape Town, Cape Town, South Africa\\
$^{111}$ University of Derby, Derby, United Kingdom\\
$^{112}$ University of Houston, Houston, Texas, United States\\
$^{113}$ University of Jyv\"{a}skyl\"{a}, Jyv\"{a}skyl\"{a}, Finland\\
$^{114}$ University of Kansas, Lawrence, Kansas, United States\\
$^{115}$ University of Liverpool, Liverpool, United Kingdom\\
$^{116}$ University of Science and Technology of China, Hefei, China\\
$^{117}$ University of Silesia in Katowice, Katowice, Poland\\
$^{118}$ University of South-Eastern Norway, Kongsberg, Norway\\
$^{119}$ University of Tennessee, Knoxville, Tennessee, United States\\
$^{120}$ University of the Witwatersrand, Johannesburg, South Africa\\
$^{121}$ University of Tokyo, Tokyo, Japan\\
$^{122}$ University of Tsukuba, Tsukuba, Japan\\
$^{123}$ Universit\"{a}t M\"{u}nster, Institut f\"{u}r Kernphysik, M\"{u}nster, Germany\\
$^{124}$ Universit\'{e} Clermont Auvergne, CNRS/IN2P3, LPC, Clermont-Ferrand, France\\
$^{125}$ Universit\'{e} de Lyon, CNRS/IN2P3, Institut de Physique des 2 Infinis de Lyon, Lyon, France\\
$^{126}$ Universit\'{e} de Strasbourg, CNRS, IPHC UMR 7178, F-67000 Strasbourg, France, Strasbourg, France\\
$^{127}$ Universit\'{e} Paris-Saclay, Centre d'Etudes de Saclay (CEA), IRFU, D\'{e}partment de Physique Nucl\'{e}aire (DPhN), Saclay, France\\
$^{128}$ Universit\'{e}  Paris-Saclay, CNRS/IN2P3, IJCLab, Orsay, France\\
$^{129}$ Universit\`{a} degli Studi di Foggia, Foggia, Italy\\
$^{130}$ Universit\`{a} del Piemonte Orientale, Vercelli, Italy\\
$^{131}$ Universit\`{a} di Brescia, Brescia, Italy\\
$^{132}$ Variable Energy Cyclotron Centre, Homi Bhabha National Institute, Kolkata, India\\
$^{133}$ Warsaw University of Technology, Warsaw, Poland\\
$^{134}$ Wayne State University, Detroit, Michigan, United States\\
$^{135}$ Yale University, New Haven, Connecticut, United States\\
$^{136}$ Yildiz Technical University, Istanbul, Turkey\\
$^{137}$ Yonsei University, Seoul, Republic of Korea\\
$^{138}$ Affiliated with an institute formerly covered by a cooperation agreement with CERN\\
$^{139}$ Affiliated with an international laboratory covered by a cooperation agreement with CERN.\\

\end{flushleft} 